\documentclass[reprint,superscriptaddress,amsmath,amssymb,aps,nofootinbib]{revtex4-1}
\usepackage{graphicx}
\usepackage[colorlinks,linkcolor=blue,citecolor=blue,urlcolor = blue]{hyperref}
\usepackage{mathtools}

\begin{document}
\title{Analytical Expression for Fracture Profile in Viscoelastic Crack Propagation}
\author{Hokuto Nagatakiya}
\thanks{These authors contributed equally: H. Nagatakiya, N. Sakumichi}
\affiliation{Graduate School of Engineering Science, The University of Osaka, 1-3 Machikaneyama, Toyonaka, Osaka 560-8531, Japan}
\author{Naoyuki Sakumichi}
\email[Corresponding author: ]{sakumichi@gel.t.u-tokyo.ac.jp}
\affiliation{Faculty of Social Informatics, ZEN University, 3-12-11 Shinjuku, Zushi, Kanagawa 249-0007, Japan}
\affiliation{Department of Chemistry and Biotechnology, The University of Tokyo, 7-3-1 Hongo, Bunkyo-ku, Tokyo 113-8656, Japan}
\author{Shunsuke Kobayashi}
\affiliation{Graduate School of Engineering Science, The University of Osaka, 1-3 Machikaneyama, Toyonaka, Osaka 560-8531, Japan}
\author{Ryuichi Tarumi}
\email[Corresponding author: ]{tarumi.ryuichi.es@osaka-u.ac.jp}
\affiliation{Graduate School of Engineering Science, The University of Osaka, 1-3 Machikaneyama, Toyonaka, Osaka 560-8531, Japan}

\date{\today}

\begin{abstract}
We derive an analytical expression for the strain field during steady-state crack propagation in viscoelastic solids described by the standard linear solid (Zener) model.
This expression reveals three regions in the fracture profile and in the strain field ahead of the crack tip, each distinguished by power-law exponents that evolve with distance from the crack tip.
These features explain the experimentally observed crack-tip sharpening in rubbers and gels as the crack-propagation velocity increases, often associated with catastrophic failure triggered by a velocity jump.
Furthermore, we establish de Gennes' viscoelastic trumpet on a continuum-mechanical foundation, previously based only on a scaling argument.
\end{abstract}

\maketitle

\textit{Introduction}.\,---
Viscoelastic rubberlike materials are used in diverse applications, including industry, medical instruments, and food products.
Understanding their fracture characteristics~\cite{Persson2005b,Knauss2015,Creton2016,Long2021} possesses both scientific and practical significance.
Rubbers~\cite{Kadir1981,Tsunoda2000,Carbone2005,Morishita2016,Sakumichi2017,Kubo2017,Kubo2021} and gels~\cite{Murai2019,Liu2019,Zhang2022} can fail catastrophically due to abrupt increases in crack-propagation velocity; such velocity jumps are strongly correlated with a rise in crack-tip sharpness.
This sharpening cannot be explained by the linear elastic fracture mechanics (LEFM)~\cite{Freund1998}, which predicts a parabolic profile for the surface of static cracks.
Experiments on rubbers~\cite{Morishita2016,Morishita2017,Mai2020} and gels~\cite{Livne2008,Bouchbinder2008,Ariel2010,Chen2017,Kolvin2018,Meng2023,Qi2024} report that increased crack-propagation velocity leads to a sharper, non-parabolic profile near the crack tip.

The weakly nonlinear theory of dynamic fracture~\cite{Bouchbinder2008}, an extension of LEFM that accommodates second-order nonlinearity but omits viscoelasticity, may explain this sharpening in some experiments~\cite{Livne2008,Ariel2010,Morishita2016}.
Nevertheless, it fails to explain the tip sharpening in which cracks propagate rapidly in filler-reinforced synthetic rubber~\cite{Morishita2016}.
This highlights the need to include viscoelasticity, which has not yet been fully incorporated into continuum fracture mechanics.

A pioneering fracture theory that incorporates viscoelasticity is de Gennes' ``viscoelastic trumpet''~\cite{deGennes1996,deGennes1997}, which adopts a standard linear solid model characterized by the frequency ($\omega$) dependence of the complex modulus
\begin{equation}
    \mu(\omega)=\mu_0 +(\mu_\infty - \mu_0)\frac{i\omega\tau}{1+i\omega\tau}.
    \label{eq:complex_modulus}
\end{equation}
Here, $\tau$ is the relaxation time, and $\mu_0$ and $\mu_{\infty}$ are the low- and high-frequency elastic moduli, respectively, with the ratio $\lambda\equiv\mu_\infty/\mu_0\sim 10^2$--$10^3$~\cite{Persson2005b,deGennes1996,deGennes1997}.
Rather than adopting the traditional fracture-mechanics approach that directly addresses continuum mechanics problems, de Gennes applied scaling analysis and energy balance.
Through this, he conjectured a trumpet-shaped fracture profile comprising three separate regions, each distinguished by characteristic power-law exponents.

However, in the absence of direct solutions in continuum mechanics, it is unclear how the viscoelastic trumpet is related to traditional fracture mechanics.
Efforts to validate the theory, whether by finite element methods (FEM) or experiments~\cite{Saulnier2004,Tabuteau2011}, have been limited, leaving a decisive confirmation unresolved.

In this Letter, we investigate how viscoelasticity leads to crack-tip sharpening as crack-propagation velocity increases.
We use the traditional fracture-mechanics approach, which seeks to explain all observed phenomena directly from the first principles of continuum mechanics.
Following de Gennes~\cite{deGennes1996,deGennes1997}, we simplify the problem by introducing certain assumptions.
We derive an analytical expression for the strain field and fracture profile during steady-state crack propagation, identifying three distinct regions within the fracture profile with different power-law exponents.
Notably, these exponents are consistent with those of the viscoelastic trumpet.
Our analytical results validate the viscoelastic trumpet and elucidate its position in continuum fracture mechanics.
They also illuminate the origin of crack-tip sharpening, which prior studies~\cite{Livne2008,Bouchbinder2008,Ariel2010,Morishita2016} attributed to the nonlinear stress--strain relationship; our analysis reveals that, even in the linear regime, viscoelasticity can amplify sharpening at elevated crack-propagation velocities.\\

\begin{figure}[t!]
    \centering
    \includegraphics[width=\linewidth]{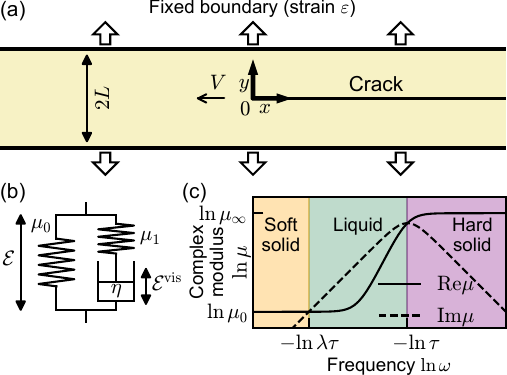}
    \caption{
    Setup of viscoelastic crack propagation.
    (a) Two-dimensional viscoelastic sheet with a semi-infinite linear crack at its center, subjected to a strain $\varepsilon$ due to fixed boundary conditions.
    The crack propagates at a constant velocity $V$ in the $-x$ direction.
    (b) Zener model: two springs with shear moduli $\mu_0$
    and $\mu_1=\mu_\infty-\mu_0$, interconnected with a dashpot with viscosity $\eta$.
    The strains of the left spring and dashpot are $\mathcal{E}_{ij}$
    and $\mathcal{E}_{ij}^{\mathrm{vis}}$, respectively.
    (c) Complex modulus $\mu(\omega)$ of the Zener model [Eq.~(\ref{eq:complex_modulus})] characterizing three types of dynamic response.
    The relaxation times are $\tau\equiv\eta/\mu_1$ and $\lambda\tau$, where $\lambda\equiv\mu_\infty/\mu_0$.
    }
    \label{fig:1}
\end{figure}

\textit{Setup}.\,---
As shown in Fig.~\ref{fig:1}(a), we consider a two-dimensional viscoelastic sheet of height $2L$ and infinite width.
A semi-infinite linear crack is located at its center, aligned with the $x$ axis in the undeformed state.
The sheet is subjected to a strain $\varepsilon$ induced by a fixed boundary condition, which causes the crack to propagate in the $-x$ direction at a constant velocity $V$.
By focusing on steady-state crack propagation, all fields at position $(x,y)$ and time $t$ depend only on the comoving coordinate $(x+Vt,y)$, allowing us to analyze the fields at time $t=0$ with the crack tip located at $x=0$.
The fixed boundary and stress-free boundary conditions are~\cite{Knauss1966,Barber1989}
\begin{align}
\begin{cases}
    u_x(x,\pm L)=0 \quad
    &(\mathrm{for}\ -\infty< x<\infty) \\
    u_y(x,\pm L)=\pm\varepsilon L
    &(\mathrm{for}\ -\infty< x<\infty) \\
    \sigma_{xy} (x,\pm 0)=\sigma_{yy}(x,\pm 0)= 0 &(\mathrm{for}\qquad 0<x<\infty),
\end{cases}\label{eq:bc}
\end{align}
where $u_i=u_i(x,y)$ is the displacement field and $\sigma_{ij}=\sigma_{ij}(x,y)$ ($i,j=x,y$) is the Cauchy stress tensor.
For $x>0$, the coordinates $(x,\pm 0)$ correspond to the upper and lower crack surfaces, respectively
\footnote{Our setup does not explicitly treat the microscopic fracture process zone at the crack tip, but instead employs a coarse-grained description valid above the process zone scale, as justified in Ref.~\cite{Qi2024}.}.

Our main objective is to determine the fracture profile (i.e., the displacement field on the crack surface) $\mathcal{U}(x)\equiv u_y(x,+0)$ for $x>0$, assuming the symmetry $u_y(x,+0)=-u_y(x,-0)$.
Following Okumura and de Gennes~\cite{Okumura2001}, we neglect $u_x(x,y)$ as it has only a minor influence on $\mathcal{U}(x)$, and thus assume
\begin{equation}
    u_x(x,y)=0,
\label{eq:ux_is_0}
\end{equation}
for $-\infty< x<\infty$ and $-L\leq y \leq L$.

In continuum mechanics, the momentum balance equation in the absence of inertia~\cite{Freund1998} reads
\begin{equation}
    0 = \sum_{j=x,y} \partial_j \sigma_{ji} \qquad
    (\mathrm{for}\ i=x,y),
    \label{eq:momentum_balance}
\end{equation}
which is a valid approximation when $V \ll C_s$, where $V$ and $C_s$ are the crack-propagation and shear wave velocities, respectively~\cite{Persson2005b,deGennes1996,deGennes1997}.

To describe viscoelasticity, we employ the Zener model~\cite{Zener1948}, shown in Fig.~\ref{fig:1}(b).
Its complex modulus $\mu(\omega)$, given by Eq.~(\ref{eq:complex_modulus}), captures the three types of dynamic response of viscoelastic solids: soft solid, liquid, and hard solid~\cite{deGennes1996,deGennes1997}, as shown in Fig.~\ref{fig:1}(c).
The viscoelastic stress--strain relationship under plane-stress conditions is
\begin{align}
    &\sigma_{ij} \equiv
    2\mu_0 \mathcal{E}_{ij} +\frac{2 \mu_0 \nu}{1-\nu} \delta_{ij}\sum_{k=x,y} \mathcal{E}_{kk} 
    + \sigma_{ij}^{\mathrm{vis}},\label{eq:stress_strain}\\
    &\sigma_{ij}^{\mathrm{vis}}\equiv 2 \eta \partial_t \mathcal{E}_{ij}^{\mathrm{vis}}
    +\frac{2 \eta \nu}{1-\nu}\delta_{ij}\sum_{k=x,y}\partial_t \mathcal{E}_{kk}^{\mathrm{vis}}\nonumber\\
    &=2\mu_1 \left(\mathcal{E}_{ij}-\mathcal{E}_{ij}^{\mathrm{vis}}\right) +\frac{2\mu_1 \nu}{1-\nu} \delta_{ij}\sum_{k=x,y} \left(\mathcal{E}_{kk}-\mathcal{E}_{kk}^{\mathrm{vis}}\right),
    \label{eq:stress_dashpot}
\end{align}
where $\mathcal{E}_{ij}\equiv(\partial_i u_j + \partial_j u_i)/2$ is the Cauchy strain, and $\sigma_{ij}^{\mathrm{vis}}$ and $\mathcal{E}_{ij}^{\mathrm{vis}}$ are the Cauchy stress and strain in the dashpot, respectively.
The details of Eqs.~(\ref{eq:stress_strain}) and (\ref{eq:stress_dashpot}) are provided Sec.~S1 in the supplemental material.
Since all fields are functions of $(x+Vt,y)$ in steady-state crack propagation, it follows that $\partial_t=V\partial_x$~\cite{Bouchbinder2008}, such as
\begin{equation}
    \partial_t\mathcal{E}_{ij}^{\mathrm{vis}}=V \partial_x \mathcal{E}_{ij}^{\mathrm{vis}}.
    \label{eq:time-derivative}
\end{equation}

\textit{Analytical solution for steady-state crack propagation}.\,---
In the setup described by Eqs.~(\ref{eq:bc})--(\ref{eq:time-derivative}), we derive the analytical solution $u_y=u_y(x,y)$ for steady-state crack propagation with a propagation velocity $V$.
This is achieved through a variable transformation \footnote{This variable transformation is analogous to that used in Graham's extended correspondence principle~\cite{Graham1968}.} that reformulates the viscoelastic steady-state crack-propagation problem into an equivalent static-crack problem where the crack does not propagate.
Accordingly, the corresponding static crack solution $u_y^0=u_y^0(x,y)$ is required.
In a static-crack problem, Eqs.~(\ref{eq:stress_strain}) and (\ref{eq:stress_dashpot}) are reduced to
\begin{equation}
    \sigma_{ij}=2\mu_0 \mathcal{E}_{ij} +\frac{2 \mu_0 \nu}{1-\nu} \delta_{ij}\sum_{k=x,y} \mathcal{E}_{kk},
    \label{eq:static_stress}
\end{equation}
because $\sigma_{ij}^{\mathrm{vis}}=0$.
This static-crack problem has been solved~\cite{Okumura2001} as follows:

\textbf{Lemma.} (Static crack)
Given the boundary conditions in Eq.~(\ref{eq:bc}) with Eq.~(\ref{eq:ux_is_0}), the displacement field satisfying Eq.~(\ref{eq:static_stress}) is given by $u_y(x,y)=u_y^0(x,y)$, where
\begin{equation}
    u_y^0 (x,y) \equiv \frac{2\varepsilon L}{\pi}\mathrm{Im}
    \left[
    \ln
    \left(
    e^{-\pi z^*/(2L)} + \sqrt{e^{-\pi z^*/L}-1}
    \right)
    \right].
     \label{eq:static_solution}
\end{equation}
Here, $z^*\equiv x\sqrt{2/(1-\nu)}-yi$ is the complex conjugate of $z\equiv  x\sqrt{2/(1-\nu)} + yi$.
In Eq.~(\ref{eq:static_solution}), the branch of the logarithmic function is selected such that $\log z$ has a zero argument for real positive $z$, and the square root is defined such that $z^{1/2}=i$ for $z=-1$.

By employing the static crack solution in Eq.~(\ref{eq:static_solution}), we obtain the solution for steady-state crack propagation as follows:

\textbf{Theorem.} (Steady-state crack propagation)
Given the boundary conditions in Eq.~(\ref{eq:bc}) with Eq.~(\ref{eq:ux_is_0}), the displacement field satisfying Eqs.~(\ref{eq:momentum_balance})--(\ref{eq:time-derivative}) is given by
\begin{equation}
    \begin{split}
    u_y (x,y) &=
    u_y^0 (x,y) 
     - \left(1 - \frac{1}{\lambda} \right)\\
    & \times\int_{-\infty}^x \frac{\partial u_y^0(\xi,y)}{\partial \xi} 
    \exp \left(-\frac{x-\xi}{\lambda \tau V}\right) d \xi,
    \label{eq:propagating_solution}
    \end{split}
\end{equation}
where $\lambda\equiv\mu_\infty/\mu_0$ and $\tau\equiv\eta/\mu_1$.
Here, $u_y^0 (x,y)$ is given by Eq.~(\ref{eq:static_solution}).

A proof of the theorem is provided in Appendix~A.
As $V\to 0$, the integrand in Eq.~(\ref{eq:propagating_solution}) vanishes, resulting in $u_y(x,y)$ converging to $u_y^0(x,y)$ in Eq.~(\ref{eq:static_solution}).\\

\begin{figure}[t!]
    \centering
    \includegraphics[width=\linewidth]{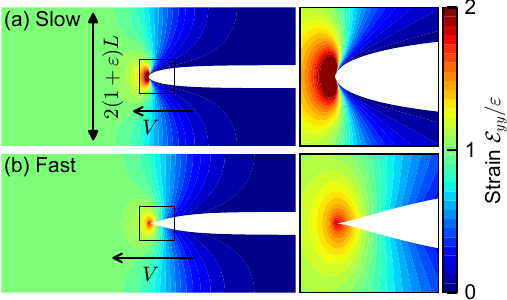}
    \caption{
    Strain field $\mathcal{E}_{yy}$ and fracture profile $\mathcal{U}(x)$ for (a) slow ($\mathcal{V}=10^{-2}$) and (b) fast ($\mathcal{V}=1$) crack propagations at a constant velocity $V$ for $\lambda=10^3$, based on the analytical expression provided in the theorem.
    }
    \label{fig:2}
\end{figure}

\textit{Analytical expression for fracture profile}.\,---
Before investigating the fracture profile $\mathcal{U}(x)\equiv u_y(x,+0)=-u_y(x,-0)$ for $x>0$ in steady-state crack propagation in Eq.~(\ref{eq:propagating_solution}), it is instructive to first confirm that the fracture profile $\mathcal{U}^0(x)\equiv u_y^0(x,+0)$ in the static crack solution in Eq.~(\ref{eq:static_solution}) exhibits a parabolic contour near the crack tip.
In the limit $y\to+0$, Eq.~(\ref{eq:static_solution}) yields~\cite{Okumura2001}
\begin{equation}
    \mathcal{U}^0(x)=\frac{2\varepsilon L}{\pi} \arctan\left[\sqrt{e^{\sqrt{2/(1-\nu)}\pi x/L}-1}\,\right]
\end{equation}
for $x>0$.
By introducing the dimensionless coordinate $X\equiv\sqrt{2/(1-\nu)} \pi x/L$, we deduce that $\mathcal{U}^0(x)=(2\varepsilon L/\pi)\sqrt{X} + O(X^{3/2})$ for small $X$, corroborating the anticipated parabolic contour near the crack tip.

To determine $\mathcal{U}(x)$, we introduce the dimensionless velocity $\mathcal{V}\equiv\sqrt{2/(1-\nu)}\pi\lambda \tau V /L$.
In the limit $y\to+0$, Eq.~(\ref{eq:propagating_solution}) yields
\begin{equation}
    \mathcal{U}(x) = \frac{2\varepsilon L}{\pi}
    \left[
    \arctan
    \left(\sqrt{e^{X}-1} \right)
    - \left(1-\frac{1}{\lambda}\right)
    H_\mathcal{V}(X)
    \right],
    \label{eq:fracture_profile}
\end{equation}
where
\begin{equation}
    \begin{split}
        &H_\mathcal{V}(X)
    \equiv
    \frac{1}{2}
    \int_0^X
    \frac{e^{-(X-\Xi)/\mathcal{V}}}{\sqrt{e^{\Xi}-1}}d\,\Xi \\
    &=e^{-X/\mathcal{V}}\sqrt{e^X-1}\, _2F_1\left(\frac{1}{2},1-\frac{1}{\mathcal{V}};\frac{3}{2};1-e^X\right).
    \end{split}
\end{equation}
Here, $\,_2 F_1(a,b;c;z)$ is the hypergeometric function~\cite{Abramowitz1965}.
This result implies that a change in the Poisson ratio $\nu$ simply rescales $X$ and $\mathcal{V}$ for $\mathcal{U}(x)$.

Based on Eq.~(\ref{eq:propagating_solution}), Fig.~\ref{fig:2} shows $\mathcal{E}_{yy}(x,y)$ and $\mathcal{U}(x)$ for slow and fast crack propagations using an arbitrary-precision library~\cite{mpmath} to avoid catastrophic cancellation.
The fracture profile exhibits a parabolic contour at lower $\mathcal{V}$, transitioning to a sharper contour as $\mathcal{V}$ increases.
These findings are consistent with the experimental observations reported for rubbers~\cite{Morishita2016,Morishita2017,Mai2020} and gels~\cite{Livne2008,Bouchbinder2008,Ariel2010,Chen2017,Kolvin2018,Meng2023,Qi2024} explained in the introduction.
\\

\begin{figure}[t!]
    \centering
    \includegraphics[width=\linewidth]{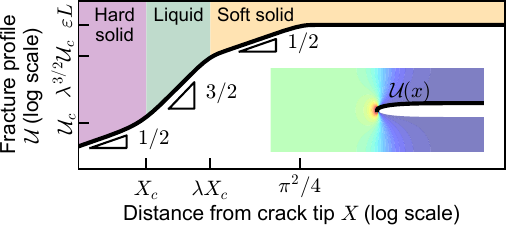}
    \caption{
    Fracture profile $\mathcal{U}(x)$ [Eq.~(\ref{eq:fracture_profile})] on a log-log scale, revealing three regions characterized by power-law exponents $1/2$, $3/2$, and $1/2$.
    The ticks on the horizontal axis indicate the crossover points, 
    $X_c$, $\lambda X_c$, and $\pi^2/4$, while the ticks on the vertical axis indicate the corresponding points of the respective asymptotic power laws, $\mathcal{U}_c$, $\lambda^{3/2} \mathcal{U}_c$, and $\varepsilon L$.
    The condition $\lambda X_c <\pi^2/4$ (equivalent to $\lambda x_c <\pi L \sqrt{1-\nu}/ 4 \sqrt{2}$) is necessary for the distinct manifestation of these regions.
    The inset shows the same profile on a linear scale, with colors corresponding to the strain field shown in Fig.~\ref{fig:2}.
    }
    \label{fig:3}
\end{figure}

\begin{figure}[t!]
    \centering
    \includegraphics[width=\linewidth]{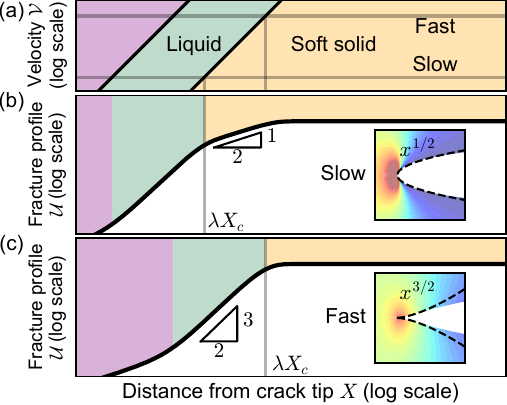}
    \caption{
    Fracture profiles $\mathcal{U}$ during slow ($\mathcal{V}=10^{-2}$) and fast ($\mathcal{V}=1$) crack propagation.
    (a) Viscoelastic states on the crack surface as a function of the distance $x$ from the crack tip, varying with $V$, while gray horizontal lines correspond to the fast and slow crack propagation cases shown in panels (b) and (c).
    Gray vertical lines indicate $\lambda X_c$ in panels (b) and (c).
    (b) At low $V$, a smaller $\lambda X_c$ leads to the dominance of the $1/2$ power-law region, resulting in a parabolic crack-tip profile.
    (c) At high $\mathcal{V}$, a larger $\lambda X_c$ extends the $3/2$ power-law region, resulting in a sharper crack-tip profile.
    }
    \label{fig:4}
\end{figure}

\textit{Origin of crack-tip sharpening and viscoelastic trumpet}.\,---
To elucidate the origin of crack-tip sharpening as $\mathcal{V}$ increases, we expand Eq.~(\ref{eq:fracture_profile}) in powers of $\sqrt{X}$ as
\begin{align*}
    \arctan\left(
    \sqrt{e^{X}-1}
    \right)
    =\sqrt{X}-\frac{X^{3/2}}{12}&+O\left(X^{5/2}\right),\\
   H_\mathcal{V}(X)=
    \sqrt{X}
   -\frac{(\mathcal{V}+8)X^{3/2}}{12 \mathcal{V}}   
   &+O\left(X^{5/2}\right).
\end{align*}
These expansions yield the asymptotic behavior of $\mathcal{U}(x)$:
\begin{equation}
    \mathcal{U}(x)\approx
    \frac{2\varepsilon L}{\pi}
    \begin{dcases}
        \frac{1}{\lambda}\sqrt{X}& (\mathrm{for}\,\, 0<X<X_c)\\
        \frac{2(\lambda -1)}{3\lambda \mathcal{V}}X^{3/2}
        & (\mathrm{for}\,\, X_c<X<\lambda X_c)\\
        \sqrt{X}& (\mathrm{for}\,\, \lambda X_c<X<\frac{\pi^2}{4}),
    \end{dcases}
\label{eq:asymptotic_U}
\end{equation}
where $X_c \equiv 3\mathcal{V}/[2(\lambda -1)]$ and $\lambda X_c$ delineate the crossover points among these regimes.
Throughout this Letter, we assume $\lambda X_c<\pi^2/4$, corresponding to $\mathcal{V}< \pi^2 (1-\lambda^{-1})/6 \approx 1.6$, to ensure that all three regions emerge before the sheet is fully relaxed [$\mathcal{U}(x)\approx\varepsilon L$] at $X \approx \pi^2/4$.
Figure~\ref{fig:3} shows $\mathcal{U}(x)$, revealing distinct power-law exponents for each regime in Eq.~(\ref{eq:asymptotic_U}): $1/2$ near the crack tip, $3/2$ beyond $X_c$ up to $\lambda X_c$, and $1/2$ again beyond $\lambda X_c$ as $H_\mathcal{V}(X)$ becomes negligible.
Since $X$ is proportional to $x$, these exponents hold in either coordinate.

The exponents ($1/2$, $3/2$, and $1/2$) and the crossover points separating each region of $\mathcal{U}(x)$ [$x_c\equiv \sqrt{1-\nu} X_c L / (\sqrt{2}\pi)=3 \lambda \tau V/[2 (\lambda -1)] \sim \tau V$ and $\lambda x_c \sim \lambda \tau V$ for $\lambda \gg 1$] in Eq.~(\ref{eq:asymptotic_U}) are consistent with those of the viscoelastic trumpet~\cite{deGennes1996,deGennes1997,Saulnier2004,hui2022}, which assumes $\lambda \gg 1$.
Notably, Eq.~(\ref{eq:asymptotic_U}) extends beyond the viscoelastic trumpet, by providing the prefactors of both the power law and the crossover points.
Although both the viscoelastic trumpet and Eq.~(\ref{eq:asymptotic_U}) rely on parameter constraints, Eq.~(\ref{eq:fracture_profile}), from which Eq.~(\ref{eq:asymptotic_U}) is derived, holds more generally.
Consequently, Eq.~(\ref{eq:fracture_profile}) reproduces the profile of the viscoelastic trumpet as a limiting case and elucidates its foundation in continuum fracture mechanics.

Figure~\ref{fig:4}(a) shows how the viscoelastic state on the crack surface evolves with the distance $x$ from the crack tip, which varies according to the crack-propagation velocity $V$.
The black solid lines represent $X_c \,(\propto \mathcal{V})$ and $\lambda X_c \,(\propto \mathcal{V})$, indicating the crossover points between the soft-solid, liquid, and hard-solid regimes.
The soft and hard solids exhibit power-law exponents of $1/2$, yielding parabolic fracture profiles, whereas the liquid exhibits a power-law exponent of $3/2$, resulting in a sharp profile.
At lower $V$ [Fig.~\ref{fig:4}(b)], the dominance of the $1/2$ power-law exponent and the small magnitudes of $\lambda X_c$ lead to nearly parabolic profiles.
In contrast, at higher $V$ [Fig.~\ref{fig:4}(c)], the increasing $\lambda X_c$ expands the liquid region, where the prevailing $3/2$ power-law exponent leads to progressive tip sharpening (note the logarithmic scale).
Thus, our analytical expression elucidates the origin of the velocity-dependent crack-tip sharpening observed in rubbers~\cite{Morishita2016,Morishita2017,Mai2020} and gels~\cite{Livne2008,Bouchbinder2008,Ariel2010,Chen2017,Kolvin2018,Meng2023,Qi2024}, suggesting that crack-tip sharpening results from high velocity, rather than the cause of velocity jump.

To further demonstrate the relevance of our analytical solution, we compare it with fracture profiles experimentally observed in carbon-black-filled rubbers~\cite{Morishita2016}, which show systematic deviations $\delta$ from a parabolic shape.
Our solution predicts $\delta$ as $\lambda x_c \approx (3/2) \lambda \tau V$, corresponding to the soft-solid--liquid crossover point (Figs.~\ref{fig:3} and \ref{fig:4}).
We focus on crack propagation at $V \approx 0.1 C_s$ (where $C_s$ is the shear wave velocity), where inertia is negligible but $\delta$ is measurable.
Using the reported relaxation time $\lambda \tau$~\cite{Morishita2016}, we find that the measured $\delta$ is consistent in magnitude with our prediction $(3/2) \lambda \tau V$.
Moreover, our analytical solution reproduces the experimentally observed increase of $\delta$ with velocity.
(See Sec.~S3 in the Supplemental Material for details.) \\

\begin{figure}[t!]
    \centering
    \includegraphics[width=\linewidth]{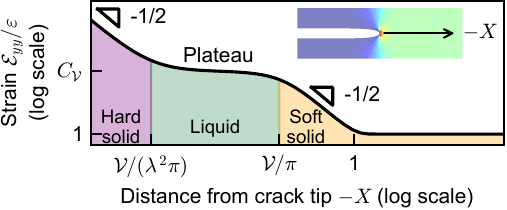}
    \caption{Strain field $\mathcal{E}_{yy}(x,0)$ ahead of the crack tip ($X<0$) as a function of the normalized distance from the crack tip $-X$ for $\lambda \gg 1$.
    The strain field exhibits two power-law regions, $\mathcal{E}_{yy}(x,0) \sim (-X) ^{-1/2}$ for $-X <\mathcal{V}/(\lambda^2\pi)$ and $\mathcal{V}/\pi<-X<1$, and a plateau for $\mathcal{V}/(\lambda^2\pi)<-X<\mathcal{V}/\pi$.
    These regions correspond to the hard-solid, liquid, and soft-solid regimes in the complex modulus [Eq.~(\ref{eq:complex_modulus})].
    }
    \label{fig:5}
\end{figure}
\textit{Strain field ahead of the crack tip}.\,---
Using Eq.~(\ref{eq:propagating_solution}), we illustrate the strain field $\mathcal{E}_{yy}(x,y) \equiv \partial_y u_y(x,y)$ ahead of the crack tip ($x<0$) along the $x$ axis ($y=0$) in Fig.~\ref{fig:5}.
Here, we assume $\lambda \gg 1$, as typical for rubberlike materials~\cite{Persson2005b,deGennes1996,deGennes1997}.
Figure~\ref{fig:5} shows that $\mathcal{E}_{yy}(x,0)$ exhibits two distinct power-law regions, $\sim (-x)^{-1/2}$, separated by an intermediate plateau, $\sim (-x)^{0}$.
These correspond to hard-solid, liquid, and soft-solid regimes in the viscoelastic response.
They are manifested in the asymptotic form
\begin{equation}
\frac{\mathcal{E}_{yy}(x,0)}{\varepsilon} \approx
\begin{dcases}
    \frac{1}{\lambda\sqrt{-X}} 
    &\, (\mathrm{for} \,\, 0<-X < \frac{1}{\lambda^2C_\mathcal{V}^2}) \\
    C_\mathcal{V}
    &\, (\mathrm{for}\,\, \frac{1}{\lambda^2C_\mathcal{V}^2} < -X < \frac{1}{C_\mathcal{V}^2}) \\
    \frac{1}{\sqrt{-X}} 
    &\, (\mathrm{for}\,\, \frac{1}{C_\mathcal{V}^2} < -X <1) \\
    1  
    &\, (\mathrm{for} \,\, 1< -X),
\end{dcases}
\label{eq:E_tip}
\end{equation}
where
\begin{equation}
C_\mathcal{V} \equiv
    \frac{\sqrt{\pi} \, \Gamma\left(\frac{1}{\mathcal{V}}+1\right)}{\Gamma\left(\frac{1}{\mathcal{V}}+\frac{1}{2}\right)}
    = \sqrt{\frac{\pi}{\mathcal{V}}}
    + \frac{\sqrt{\pi\mathcal{V}}}{8}
     + O(\mathcal{V}^{3/2}),
     \label{eq:Eyy_plateau}
\end{equation}
and $\Gamma(z)$ is the gamma function.
The derivation of Eq.~(\ref{eq:E_tip}) is provided in Appendix~B.

For $\mathcal{V} \ll 1$, the crossover points in Eq.~(\ref{eq:E_tip}) are $-X = 1/C_\mathcal{V}^2 \approx \mathcal{V}/\pi$ and $-X = 1/(\lambda^2 C_\mathcal{V}^2) \approx \mathcal{V}/(\pi \lambda^2)$.
The former corresponds to the soft-solid--liquid crossover and gives $-x \approx \lambda \tau V / \pi \sim \lambda \tau V$, which coincides with the characteristic length scale in the fracture profile [Eq.~(\ref{eq:asymptotic_U})].
The latter corresponds to the liquid--hard-solid crossover and gives $-x \approx \tau V / (\pi \lambda) \sim \tau V/\lambda$, which is more localized by a factor of $\lambda^{-1}$ relative to the scale of the fracture profile.

Equation~(\ref{eq:E_tip}) is consistent with strain-field measurements in carbon-black reinforced styrene--butadiene rubber (SBR)~\cite{Mai2020}, which exhibit a plateau near the crack tip followed by a power-law decay. 
Focusing on the low-velocity case ($V \approx 0.03 C_s \approx 0.6$~m/s), where inertia is negligible, we estimate the soft-solid--liquid crossover length as $\lambda \tau V/\pi$ using a representative relaxation time $\lambda \tau \approx 10^{-3}$~s.
The resulting value ($\approx 0.2$~mm) agrees with the experimentally observed plateau width, suggesting that the strain plateau arises from viscoelasticity rather than cohesive-zone effects.
(See Sec.~S4 in the Supplemental Material for details.)\\

\textit{Conclusion}.\,---
We have derived an analytical expression [Eq.~(\ref{eq:fracture_profile})] for the fracture profile during steady-state crack propagation in viscoelastic solids, utilizing the standard linear solid model.
The resulting fracture profile exhibits three distinct regions [Eq.~(\ref{eq:asymptotic_U}) and Fig.~\ref{fig:3}], each characterized by a power-law exponent, confirming de Gennes' viscoelastic trumpet~\cite{deGennes1996,deGennes1997}.
This expression further provides the power-law prefactors and crossover locations beyond the original scaling argument.
We also analyzed the strain field ahead of the crack tip [Eq.~(\ref{eq:E_tip}) and Fig.~\ref{fig:5}], revealing a plateau between two power-law regimes with $(-x)^{-1/2}$, a signature of viscoelasticity.
The predictions capture features experimentally observed in fracture profiles~\cite{Morishita2016} and strain fields~\cite{Mai2020} in filled rubbers.
These findings bridge a long-standing gap in viscoelastic fracture mechanics by linking the crack-tip sharpness to propagation velocity [Fig.~\ref{fig:4}].
By emphasizing the critical role of viscoelasticity, our research could facilitate the development of tough polymer materials by controlling crack-tip processes.

\textit{Acknowledgments}.\,---
We thank Professor~K.~Urayama and Professor~T.~Sakai for helpful experimental insights.
This work was supported by JST PRESTO Grant Number JPMJPR1997, JST FOREST Program Grant Number JPMJFR232A, and JST ERATO Grant Number JPMJER2401. 
This work was also supported by JSPS KAKENHI Grant Numbers JP22H01187, JP23K22458, and JP25K00966.\\

\renewcommand{\theequation}{A\arabic{equation}}
\setcounter{equation}{0}

\textit{Appendix A: Proof of the Theorem}.\,---
To prove Eq.~(\ref{eq:propagating_solution}), we introduce a transformation relating the intended displacement field $u_i(x,y)$ to an auxiliary field $\mathfrak{u}_i^0(x,y)$
(for $i=x,y$), defined by
\begin{align}
    \mathfrak{u}_i^0(x,y) &\equiv (1+\lambda \tau V \partial_x)u_i^{\mathrm{vis}}(x,y)
    \label{eq:def_u0},\\
    u_i^{\mathrm{vis}} (x,y)&\equiv \frac{1}{\tau V}
    \int_{-\infty}^x u_i(\xi,y) e^{-(x-\xi)/(\tau V)} d\xi.
    \label{eq:def_u_dashpot}
\end{align}
Equation~(\ref{eq:def_u_dashpot}) implies $u_i(x,y)=(1+\tau V \partial_x)u_i^{\mathrm{vis}}(x,y)$.
First, we show that $\mathfrak{u}_i^0(x,y)$ satisfies both the balance equation [Eq.~(\ref{eq:momentum_balance})] and the boundary conditions in the static crack.
Substituting Eq.~(\ref{eq:time-derivative}) into Eq.~(\ref{eq:stress_dashpot}), we obtain $\mathcal{E}_{ij}=(1+\tau V \partial_x)\mathcal{E}_{ij}^{\mathrm{vis}}$.
Combining this relationship with Eq.~(\ref{eq:def_u_dashpot}), we obtain
$\mathcal{E}_{ij}^{\mathrm{vis}}=(\partial_j u_i^{\mathrm{vis}} + \partial_i u_j^{\mathrm{vis}})/2$.
Substituting Eqs.~(\ref{eq:stress_dashpot}) and (\ref{eq:time-derivative}) into Eq.~(\ref{eq:stress_strain}), we obtain
$\sigma_{ij}=2\mu_0 (1+\lambda \tau V\partial_x)\mathcal{E}_{ij}^{\mathrm{vis}}$.
Combining this equation with Eq.~(\ref{eq:def_u0}), we obtain the stress--strain relationship in the static crack, Eq.~(\ref{eq:static_stress}) with $\mathcal{E}_{ij}$ replaced by $\mathcal{E}_{ij}^0\equiv (\partial_j \mathfrak{u}_i^0+\partial_i \mathfrak{u}_j^0)/2$.
Therefore, $\mathfrak{u}_i^0(x,y)$ satisfies the balance equation in the static crack, as claimed above. 

Applying Eq.~(\ref{eq:bc}) to Eqs.~(\ref{eq:def_u0}) and (\ref{eq:def_u_dashpot}), we derive the boundary conditions for $\mathfrak{u}_i^0$ as
\begin{align}
\begin{cases}
    \mathfrak{u}_x^0(x,\pm L)=0& (\mathrm{for}\ -\infty< x <\infty)\\
    \mathfrak{u}_y^0(x,\pm L)=\pm \varepsilon L
    & (\mathrm{for}\ -\infty< x <\infty).
\end{cases}
\label{eq:bc_u0}
\end{align}
These boundary conditions for $\mathfrak{u}_i^0$ are thus the same as those for $u_i$ in Eq.~(\ref{eq:bc}); additional stress conditions are the same as those in Eq.~(\ref{eq:bc}).
Therefore, $\mathfrak{u}_i^0(x,y)$ is simply the displacement field in the static crack, as described in the Lemma.

Next, we derive $u_y(x,y)$.
Using Eqs.~(\ref{eq:ux_is_0}), (\ref{eq:def_u0}), and  (\ref{eq:def_u_dashpot}), we obtain $\mathfrak{u}_x^0(x,y)=0$.
Substituting Eq.~(\ref{eq:static_stress}) into Eq.~(\ref{eq:momentum_balance}) with $\mathcal{E}_{ij}$ replaced by $\mathcal{E}_{ij}^0$, we obtain
\begin{equation}
    \left[(1+\nu) \partial_x^2+2 \partial_y^2\right] \mathfrak{u}_y^0(x,y)=0,
   \label{eq:Laplace_equation}
\end{equation}
which has the form of the Laplace equation.
The solution of Eq.~(\ref{eq:Laplace_equation}), subject to the boundary conditions [Eq.~(\ref{eq:bc_u0}) and the stress conditions in Eq.~(\ref{eq:bc})], is $\mathfrak{u}_y^0(x,y)=u_y^0(x,y)$, as provided in Eq.~(\ref{eq:static_solution}).
To determine $u_y(x,y)$, we employ the variation of constants method.
By using Eqs.~(\ref{eq:static_solution}), (\ref{eq:def_u0}), and (\ref{eq:def_u_dashpot}), we derive $u_y(x,y)$ as given in Eq.~(\ref{eq:propagating_solution}).
\quad Q.E.D.\\

\renewcommand{\theequation}{B\arabic{equation}}
\setcounter{equation}{0}

\textit{Appendix B: Derivation of strain field ahead of the crack tip}.\,---
In this appendix, we analyze the strain field $\mathcal{E}_{yy}(x,y) \equiv \partial_y u_y(x,y)$ and derive its asymptotic form given in Eq.~(\ref{eq:E_tip}).
Using the analytical expression for $u_y(x,y)$ in Eq.~(\ref{eq:propagating_solution}), the strain field is expressed as
\begin{equation}
\begin{split}
    & \mathcal{E}_{yy}(x,y) 
    = \mathcal{E}_{yy}^0 (x,y) \\
    & - \left(1 - \frac{1}{\lambda} \right)
    \int_{-\infty}^x \frac{\partial \mathcal{E}_{yy}^0(\xi,y)}{\partial \xi} 
    \exp \left(-\frac{x-\xi}{\lambda \tau V}\right) d\xi,
    \label{eq:strain}
\end{split}
\end{equation}
where $\mathcal{E}_{yy}^0 (x,y) \equiv \partial_y u_y^0 (x,y)$ is the strain field for the static crack, and the second term represents the viscoelastic contribution arising from crack propagation. 
By integrating Eq.~(\ref{eq:strain}) by parts, we obtain
\begin{equation}
    \mathcal{E}_{yy}(x,y) 
    = \frac{1}{\lambda} \mathcal{E}_{yy}^0(x,y) + \left(1-\frac{1}{\lambda}\right)\mathcal{E}_{yy}^{\mathrm{vis}}(x,y),
\label{eq:S:E_yy}
\end{equation}
where
\begin{equation}
     \mathcal{E}_{yy}^{\mathrm{vis}}(x,y)
     =  \frac{1}{\lambda \tau V}\int_{-\infty}^x \mathcal{E}_{yy}^0(\xi,y) \exp\left(-\frac{x-\xi}{\lambda\tau V}\right) d\xi
     \label{eq:A3}
\end{equation}
is the strain field in the dashpot.
Along the line $y = 0$, Eq.~(\ref{eq:static_solution}) yields
\begin{equation}
    \frac{\mathcal{E}_{yy}^0(x,0)}{\varepsilon} =
    \begin{dcases}
        \frac{1}{\sqrt{1-e^X}} & \quad (\mathrm{for} \ x<0)\\
        0 & \quad (\mathrm{for} \ x>0),
    \end{dcases}
    \label{eq:S:E_yy0}
\end{equation}
with the dimensionless coordinate $X \equiv \sqrt{2/(1-\nu)}\pi x /L$.
Substituting Eq.~(\ref{eq:S:E_yy0}) into Eq.~(\ref{eq:A3}), we obtain the strain field in the dashpot ahead of the crack tip ($x<0$) as
\begin{equation}
\begin{split}
    & \frac{\mathcal{E}_{yy}^{\mathrm{vis}}(x,0)}{\varepsilon}
     =\frac{1}{\mathcal{V}}\int_{-\infty}^X  \frac{1}{\sqrt{1-e^\Xi}} \exp\left(-\frac{X-\Xi}{\mathcal{V}}\right) d\Xi\\
     & \qquad\quad = \sqrt{1-e^X} \,_2 F_1\left(1,\frac{1}{2}+\frac{1}{\mathcal{V}};1+\frac{1}{\mathcal{V}};e^{X}\right),
    \label{eq:S:E_yy_vis}
\end{split}
\end{equation}
with the dimensionless crack-propagation velocity $\mathcal{V} \equiv \sqrt{2/(1-\nu)} \pi \lambda \tau V/L$ and the hypergeometric function $\,_2 F_1(a,b;c;z)$.

Figure~\ref{fig:5} shows the strain field $\mathcal{E}_{yy}(x,0)$ ahead of the crack tip ($x<0$) using Eqs.~(\ref{eq:S:E_yy}), (\ref{eq:S:E_yy0}), and (\ref{eq:S:E_yy_vis}) for $\lambda \gg 1$.
The strain field exhibits two distinct power-law regimes separated by an intermediate plateau.

To clarify the asymptotic behavior near the crack tip, we expand $\mathcal{E}_{yy}^0(x,0)$ in Eq.~(\ref{eq:S:E_yy0}) for $X<0$ as
\begin{equation}
\begin{split}
    \frac{\mathcal{E}_{yy}^0(x,0)}{\varepsilon}
    & = \frac{1}{\sqrt{1-e^X}} \\
    & = \frac{1}{\sqrt{-X}} + \frac{\sqrt{-X}}{4} + O((-X)^{3/2}),
     \label{eq:E_yy0_expand}
\end{split}
\end{equation}
indicating a singularity of the form $(-X)^{-1/2}$ as $X \to -0$.
Similarly, expanding Eq.~(\ref{eq:S:E_yy_vis}) for $X<0$ yields
\begin{equation}
    \frac{\mathcal{E}_{yy}^{\mathrm{vis}}(x,0)}{\varepsilon}
    = C_\mathcal{V}
    - \frac{2\sqrt{-X}}{\mathcal{V}}
    +  O((-X)^{1}),
     \label{eq:E_yy_vis_expand}
\end{equation}
where $C_\mathcal{V}$ is defined in Eq.~(\ref{eq:Eyy_plateau}).
Equation~(\ref{eq:E_yy_vis_expand}) indicates that $\mathcal{E}_{yy}^{\mathrm{vis}}(x,0)$ remains finite as $X \to -0$.
Substituting Eqs.~(\ref{eq:E_yy0_expand}) and (\ref{eq:E_yy_vis_expand}) into Eq.~(\ref{eq:S:E_yy}), we obtain the asymptotic form of the strain field as
\begin{equation}
    \frac{\mathcal{E}_{yy}(x,0)}{\varepsilon}
    = \frac{1}{\lambda \sqrt{-X}}
    + \left(1-\frac{1}{\lambda} \right)
    C_\mathcal{V}
    + O(\sqrt{-X}).
    \label{eq:Eyy_Asym}
\end{equation}
As $X \to -0$, the first term in Eq.~(\ref{eq:Eyy_Asym}) dominates and diverges as $(-X)^{-1/2}$, corresponding to the hard-solid regime.
As $-X$ increases, the singular term in Eq.~(\ref{eq:Eyy_Asym}) rapidly decays, particularly for large $\lambda$, and $\mathcal{E}_{yy}(x,0)$ begins to exhibit a plateau governed by the second term in Eq.~(\ref{eq:Eyy_Asym}), corresponding to the viscoelastic liquid regime.
As $-X$ increases further, the second term (the viscoelastic contribution) in Eq.~(\ref{eq:strain}) becomes negligible, and the first term becomes dominant.
Thus, the static strain field $\mathcal{E}_{yy}^0(x,0) \approx 1/\sqrt{-X}$ emerges near $-X \approx 1$ [see Eq.~(\ref{eq:E_yy0_expand})], corresponding to the soft-solid regime.
These results show that the strain field ahead of the crack tip $\mathcal{E}_{yy}(x,0)$ exhibits three distinct regimes: a near-tip hard-solid regime, an intermediate liquid-like plateau, and a distant soft-solid regime, as captured by Eq.~(\ref{eq:E_tip}).

\bibliography{bibliography}

\end{document}


\title{Supplemental Material for:\\
``Analytical Expression for Fracture Profile in Viscoelastic Crack Propagation''}

\author{Hokuto Nagatakiya}
\thanks{These authors contributed equally: H. Nagatakiya, N. Sakumichi}
\affiliation{Graduate School of Engineering Science, The University of Osaka, 1-3 Machikaneyama, Toyonaka, Osaka 560-8531, Japan}
\author{Naoyuki Sakumichi}
\email[Corresponding author: ]{sakumichi@gel.t.u-tokyo.ac.jp}
\affiliation{Faculty of Social Informatics, ZEN University, 3-12-11 Shinjuku, Zushi, Kanagawa 249-0007, Japan}
\affiliation{Department of Chemistry and Biotechnology, The University of Tokyo, 7-3-1 Hongo, Bunkyo-ku, Tokyo 113-8656, Japan}

\author{Shunsuke Kobayashi}
\affiliation{Graduate School of Engineering Science, The University of Osaka, 1-3 Machikaneyama, Toyonaka, Osaka 560-8531, Japan}

\author{Ryuichi Tarumi}
\email[Corresponding author: ]{tarumi.ryuichi.es@osaka-u.ac.jp}
\affiliation{Graduate School of Engineering Science, The University of Osaka, 1-3 Machikaneyama, Toyonaka, Osaka 560-8531, Japan}

\date{\today}

\maketitle

\vspace{-1cm}

\section{Viscoelastic stress--strain relationship of the Zener model}

This section provides a detailed explanation of the viscoelastic stress--strain relationship of the Zener model [Eqs.~(5) and (6) in the main text].
For two-dimensional isotropic linear elastic solids under a plane-stress condition, the Cauchy stress ($\sigma_{ij}$) relates to the Cauchy strain $\mathcal{E}_{ij} \equiv (\partial_j u_i + \partial_i u_j)/2$ via~\cite{Landau1986theory}
\begin{equation}
    \sigma_{ij} = 2 \mu_0 \mathcal{E}_{ij} + \frac{2 \mu_0 \nu}{1-\nu} \delta_{ij}\sum_{k=x,y} \mathcal{E}_{kk},
    \label{eq:S:static_stress_nu}
\end{equation}
where $\mu_0$ is the (equilibrium) shear modulus, $\nu$ is the Poisson ratio, and $\delta_{ij}$ is the Kronecker delta.
Assuming an identical Poisson ratio for both springs (shear moduli $\mu_0$ and $\mu_1$) and the dashpot (viscous modulus $\eta$) [Fig.~1(b) in the main text], the viscoelastic stress--strain relationship of the Zener model under a plane-stress condition becomes
\begin{align}
    \sigma_{ij} & = 2\mu_0 \mathcal{E}_{ij} +\frac{2 \mu_0 \nu}{1-\nu} \delta_{ij}\sum_{k=x,y} \mathcal{E}_{kk} 
    + \sigma_{ij}^{\mathrm{vis}},\label{eq:S:stress_strain}\\
    \sigma_{ij}^{\mathrm{vis}} & = 2 \eta \partial_t \mathcal{E}_{ij}^{\mathrm{vis}}
    +\frac{2 \eta \nu}{1-\nu} \delta_{ij}\sum_{k=x,y}\partial_t \mathcal{E}_{kk}^{\mathrm{vis}}
    =2\mu_1 \left(\mathcal{E}_{ij}-\mathcal{E}_{ij}^{\mathrm{vis}}\right) +\frac{2 \mu_1 \nu}{1-\nu} \delta_{ij}\sum_{k=x,y} \left(\mathcal{E}_{kk}-\mathcal{E}_{kk}^{\mathrm{vis}}\right)
    \label{eq:S:stress_vis},
\end{align}
where $\sigma_{ij}^{\mathrm{vis}}$ and $\mathcal{E}_{ij}^{\mathrm{vis}}$ are the Cauchy stress and strain in the dashpot, respectively.
Equations~(\ref{eq:S:stress_strain}) and (\ref{eq:S:stress_vis}) reproduce Eqs.~(5) and (6) in the main text.

Substituting Eq.~(\ref{eq:S:static_stress_nu}) into the momentum balance equation $0=\sum_{j=x,y}\partial_j \sigma_{ji}$ yields coupled equations for the displacement field $u_x(x,y)$ and $u_y(x,y)$:
\begin{equation}
    \left(
    \frac{\partial^2}{\partial x^2}+\frac{1-\nu}{2}  \frac{\partial^2}{\partial y^2}
    \right)u_x
    +\frac{1+\nu}{2}  \frac{\partial^2 u_y}{\partial x \partial y}
    =0,
    \label{eq:S:momentum_balance1}
\end{equation}
\begin{equation}
    \frac{1+\nu}{2}  \frac{\partial^2 u_x}{\partial x \partial y}+
    \left(
    \frac{1-\nu}{2}  \frac{\partial^2}{\partial x^2}+\frac{\partial^2}{\partial y^2}
    \right)u_y=0.
    \label{eq:S:momentum_balance2}
\end{equation}
The coefficients in Eqs.~(\ref{eq:S:momentum_balance1}) and (\ref{eq:S:momentum_balance2}) depend on the Poisson ratio $\nu$ and are nonzero for $-1<\nu<1/2$, confirming that the assumption $u_x(x,y)=0$ [Eq.~(3) in the main text] is independent of $\nu$.

\section{Numerical Verification of Analytical Expression}

In this section, we numerically verify our analytical expression for the strain field and fracture profile [Eq.~(9) in the main text], by using the finite element method (FEM).
We consider a two-dimensional domain $\Omega$ extending from $-30L$ to $30L$ in the $x$-direction and from $-L$ to $L$ in the $y$-direction, with a straight crack extending from the origin along the positive $x$-axis.
To ensure numerical stability, the crack is represented by a very thin (nearly straight) triangular notch with angle $\theta = 3.8\times10^{-6}$~rad.
For FEM calculations, we describe the governing equations and boundary conditions [Eqs.~(2)--(6) in the main text] in weak form as the first variation of the action integral $\delta I = 0$, where
\begin{equation}
    \delta I = \int_{\Omega}dV \left(
     \sum_{i,j=x,y}\partial_j \sigma_{ji} \delta u_i
     + \sum_{k,l=x,y} \sigma_{kl}^{\mathrm{const}} \delta \mathcal{E}_{kl}^{\mathrm{vis}}
     \right),
\label{eq:S:weakform}
\end{equation}
and 
\begin{align}
    \sigma_{ij}^{\mathrm{const}} \equiv
    2 \eta V\partial_x \mathcal{E}_{ij}^{\mathrm{vis}}
    +\frac{2 \eta \nu}{1-\nu} \delta_{ij}\sum_{k=x,y}V\partial_x \mathcal{E}_{kk}^{\mathrm{vis}}
    -2\mu_1 \left(\mathcal{E}_{ij}-\mathcal{E}_{ij}^{\mathrm{vis}}\right) -\frac{2 \mu_1 \nu}{1-\nu} \delta_{ij}\sum_{k=x,y} \left(\mathcal{E}_{kk}-\mathcal{E}_{kk}^{\mathrm{vis}}\right).
\end{align}
The condition $\delta I = 0$ for arbitrary functions $\delta u_i$ and $\delta \mathcal{E}_{kl}^{\mathrm{vis}}$ is equivalent to the momentum balance equation [Eq.~(4) in the main text], the viscoelastic stress--strain relationship in the dashpot [Eq.~(6) in the main text], and the steady-state crack propagation relation $\partial_t = V \partial_x$.
Here, $\delta u_x = 0$ when assuming $u_x = 0$.
We conducted FEM simulations using the analysis software FEniCS~\cite{FEniCS}, with triangular elements generated by the meshing software Gmsh~\cite{Gmsh}.
The elements employed second-order basis functions and five-point Gaussian quadrature for numerical integration.

\begin{figure}
    \centering
    \includegraphics[width=0.84\linewidth]{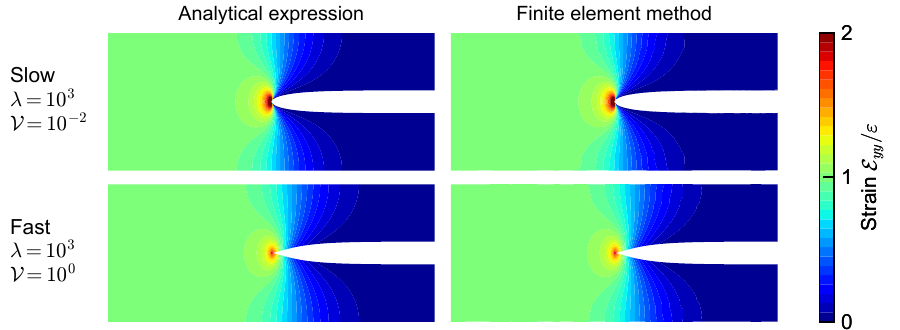}
    \caption{Comparison of strain fields $\mathcal{E}_{yy}/\varepsilon$ and fracture profiles obtained from the analytical expression [Eq.~(9) in the main text] and FEM simulations.
    The parameters used are identical to those in Fig.~2 in the main text, i.e., $\lambda = 10^3$ with $\mathcal{V}=10^{-2}$ and $\mathcal{V}=1$
    for slow and fast crack propagation, respectively.
    }
    \label{fig:Sfem_strain_ux=0}
\end{figure}

\begin{figure}
    \centering
    \includegraphics[width=0.85\linewidth]{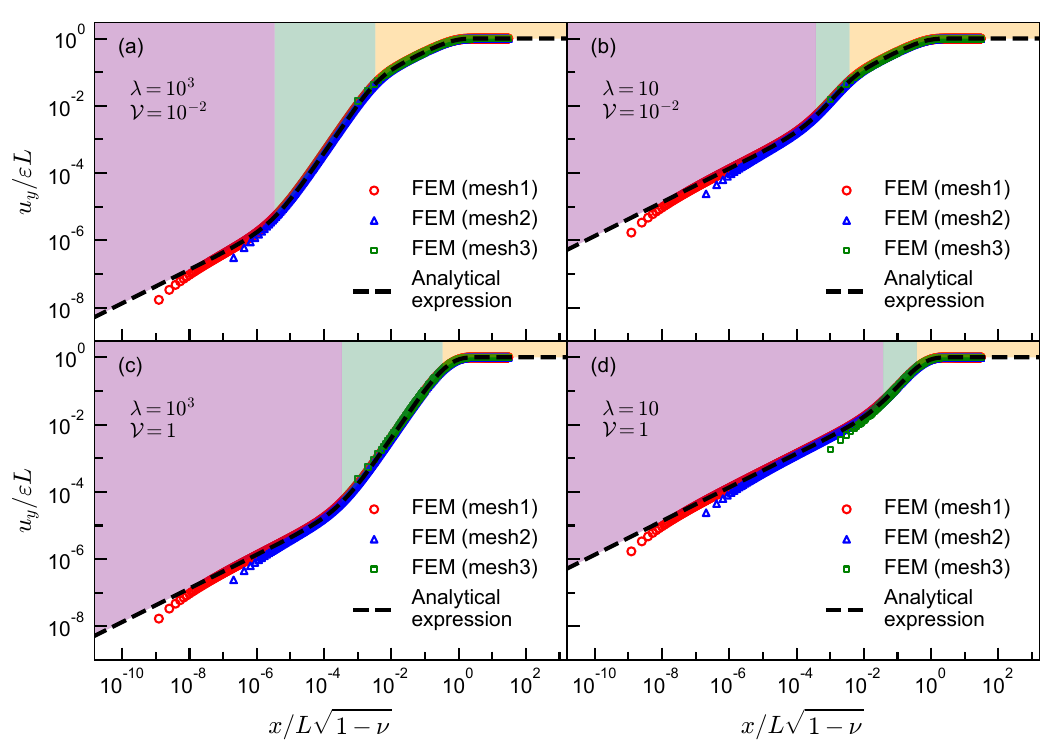}
    \caption{Comparison of analytical and simulated fracture profiles for three element sizes: $\Delta x/L \sim 10^{-9}$ (mesh 1), $10^{-7}$ (mesh 2), and $10^{-3}$ (mesh 3).
    The parameters used in (a) and (c) are identical to those in Fig.~4 (b) and (c) in the main text, respectively.
    The minimum and maximum points for $x/L$ on each mesh correspond to the element size ($\Delta x$) and geometry size ($30 L$), respectively.
    The colors at the top of each panel indicate the states of the viscoelastic solid on the crack surface: soft solid (yellow), viscous liquid (light green), and hard solid (purple).
    }
    \label{fig:S:fem_solution_compare}
\end{figure}

Figure \ref{fig:Sfem_strain_ux=0} shows the strain fields $\mathcal{E}_{yy}/\varepsilon$ and fracture profiles obtained from both the analytical expression and FEM simulations.
The analytical expression derived in the main text is in good agreement with the numerical solutions obtained through FEM, confirming the validity of the analytical expression.
The analysis employs an adaptive mesh that becomes progressively finer as it approaches the crack tip, with the mesh size at the crack tip being $\Delta x/L \sim 10^{-9}$.

To examine the influence of element size on the FEM simulations, we compare three distinct meshes with varying element sizes.
Figure~\ref{fig:S:fem_solution_compare} shows a comparison of the fracture profile from the analytical expression with those from FEM simulations, using different parameters for $\lambda$, $\mathcal{V}$, and element sizes of $\Delta x/L \sim 10^{-9}$ (mesh 1), $10^{-7}$ (mesh 2), and $10^{-3}$ (mesh 3).
The minimum and maximum points on each mesh in each panel correspond to the minimum element size ($\Delta x$) and the system size ($30 L$), respectively.
Although minor discrepancies between the FEM results and the analytical expression exist near the crack tip in a few elements, these differences converge within approximately 10 elements for each mesh.
The discrepancies may be attributed to the finite size effect, which tends to stiffen the deformation.

\section{Comparison with Experiments 1: Fracture Profile}

To demonstrate the relevance of our analytical solution, we compare it with the experimentally observed fracture profiles reported by Morishita et al.~\cite{Morishita2016}.
Their study provides a benchmark for evaluating crack-tip deviations in carbon-black-filled rubber, where the fracture profile systematically deviates from a parabolic shape as the crack-propagation velocity increases.
Figure 6 of Ref.~\cite{Morishita2016} presents experimentally observed fracture profiles for three crack-propagation velocities:
(a) a slow crack, (b) a fast crack, and (c) an even faster crack.
These fracture profiles clearly show that the crack tip becomes progressively sharper as the propagation velocity increases.
Furthermore, Fig.~7 of Ref.~\cite{Morishita2016} quantifies the relationship between the crack-propagation velocity $V$ (normalized by the shear wave velocity $C_s$) and the crack-tip deviation $\delta$ from a parabolic profile, demonstrating that $\delta$ remains small for slow cracks but increases significantly  for fast cracks.
Thus, our analytical solution successfully captures this experimentally observed increase in $\delta$ with increasing crack-propagation velocity $V$.

To further compare our theoretical predictions with experiment quantitatively, we focus on the relationship among the crack-propagation velocity $V$, the deviation $\delta$, and the storage modulus $G'(\omega)$.
Since shift factor data ($a_T$) are available only for the carbon-black (CB) volume fraction $\phi=0.14$ [Fig.~2(b) of Ref.~\cite{Morishita2016}], we restrict our analysis to this case.
Our analytical solution is derived under the assumptions that inertia is negligible and that viscoelasticity-induced crack-tip sharpening dominates over nonlinearity-induced sharpening.
This assumption is valid in the around $V/C_s \sim 10^{-3}$, corresponding to the low-velocity data in Fig.~7 of Ref.~\cite{Morishita2016}.
For $\phi=0.14$, the shear wave velocity is measured as $C_s\approx 50$~m/s [see Fig.~4(a) of Ref.~\cite{Morishita2016}], giving $V\approx 0.05$~m/s.
According to our analytical solution, the deviation $\delta$ is given by 
\begin{equation}
\delta=\frac{3}{2} \lambda \tau V,
\end{equation}
where $\lambda \tau$ is the relaxation time that defines the boundary between the soft-solid ($\sim x^{1/2}$) and liquid ($\sim x^{3/2}$) regimes. 
This relaxation time ($\lambda \tau$) corresponds to the inverse of the angular frequency at which the viscoelastic material transitions from soft-solid [$G'(\omega)>G''(\omega)$] to liquid [$G'(\omega)<G''(\omega)$] regime.
Substituting the experimental values $\delta\approx 0.5$~mm and $V\approx 0.05$~m/s into the above relation, we estimate $\lambda \tau \approx 6.7\times 10^{-3}$~s.

To validate this estimate, we examine the storage modulus $G'(\omega)$ shown in Fig.~2(a) of Ref.~\cite{Morishita2016}.
The experimental data indicate that the transition from the soft-solid to the liquid regime occurs at a characteristic angular frequency of $\omega= 1/(\lambda \tau)\approx1/(6.7\times 10^{-3})\approx 1.5\times 10^2$~$\mathrm{rad/s}$, where $a_T=1$ at a temperature of $T=25^\circ$C as shown in Fig.~2(b).
At this frequency, $G'(\omega)$ exhibits an inflection point in its frequency dependence, indicating the transition from soft-solid [$G'(\omega)>G''(\omega)$] to liquid [$G'(\omega)<G''(\omega)$] regime.
This agreement between our theoretically predicted $\lambda \tau$ and the characteristic relaxation time extracted from rheological measurements confirms that our analytical solution accurately captures viscoelastic crack-tip sharpening.
The observed quantitative consistency with the experimental data of Ref.~\cite{Morishita2016} validates our theoretical framework in describing dynamic fracture in viscoelastic materials.

\section{Comparison with Experiments 2: Strain Plateau Ahead of the Crack Tip}

To demonstrate the experimental relevance of our analytical solution, we compare it with the strain field measurements in carbon-black reinforced styrene-butadiene rubber (CB/SBR) reported by Mai et al.~\cite{Mai2020}.
Figure~3 of Ref.~\cite{Mai2020} shows the strain distribution ahead of a crack tip, exhibiting a plateau near the crack tip and a region where the strain decreases with increasing distance from the crack tip.
Our analytical solution predicts a three-regime behavior consisting of two power-law regions separated by an intermediate strain plateau [see Fig.~5 in the main text].
However, the inner power-law region (hard-solid regime) near the crack tip is extremely narrow (approximately six orders of magnitude smaller than the plateau length; that is, $\sim \lambda^{-2} \approx 10^{-6}$) and thus far below the spatial resolution ($\approx 0.2$~mm) in Ref.~\cite{Mai2020}.
As a result, only the strain plateau (liquid regime) and the outer power-law region (soft-solid regime) are experimentally observable, both of which are captured by our theory.
Since the strain plateau reported in Ref.~\cite{Mai2020} is approximately 0.6~mm, far exceeding typical polymer-network length scales, it is reasonable to attribute it to viscoelastic relaxation rather than to microscopic fracture processes such as those described by cohesive-zone models.
Below, we show that the experimentally observed plateau length is consistent with our theoretical prediction.

As shown in Fig.~5 in the main text, the crossover point between the strain plateau (liquid) and the outer power-law (soft-solid) region occurs at a dimensionless position $-X = 1/C_\mathcal{V}^2 \approx \mathcal{V}/\pi$.
Restoring the dimensionless quantities $X \equiv \sqrt{2/(1-\nu)}\pi x /L$ and $\mathcal{V} \equiv \sqrt{2/(1-\nu)}\pi \lambda \tau V /L$ to their dimensional forms, the corresponding crossover length reads
\begin{equation}
    -x \approx \frac{\lambda \tau V}{\pi},
    \label{eq:S:x_LS}
\end{equation}
where $\lambda \tau$ is the longest relaxation time in the Zener model.

To validate Eq.~(\ref{eq:S:x_LS}), we extract material parameters from the literature.
Figure~3 of Ref.~\cite{Mai2020} shows the strain distributions as a function of the distance from the crack tip, for two crack-propagation velocities: $V = 0.03 C_s$ and $V = 1.84 C_s$, where $C_s$ is the shear wave velocity.
Since our theory neglects inertia, it is applicable for  the low-velocity regime ($V \ll C_s$), and we therefore focus on $V = 0.03 C_s$.
Figure~1(c) of Ref.~\cite{Mai2020} reports $C_s \approx 20$~m/s, implying $V \approx 0.6$~m/s.
The crossover from the strain plateau to the outer power-law region occurs at $-x\approx 0.6$~mm.
Substituting $V \approx 0.6$~m/s and $-x\approx 0.6$~mm into Eq.~(\ref{eq:S:x_LS}) yields an estimate of $\lambda \tau \approx 3 \times 10^{-3}$~s.

Although Ref.~\cite{Mai2020} does not report the relaxation time or complex modulus, we refer to Fig.~11 of Ref.~\cite{Kubo2021}, which shows the storage modulus $G'(\omega)$ of SBR.
The characteristic relaxation time $\lambda \tau$ corresponds to the inverse of the crossover angular frequency $\omega_c$, which separates the low-frequency rubbery plateau and the rubbery-to-glassy transition in $G'(\omega)$.
Assuming $\lambda \tau \approx 3 \times 10^{-3}$~s, as estimated above, the corresponding angular frequency is $\omega_c = 2\pi/(\lambda \tau) \approx 2 \times 10^3$~rad/s.
This value is consistent with the observed crossover in $G'(\omega)$ around $10^3$--$10^4$~rad/s in Fig.~11 of Ref.~\cite{Kubo2021}.
Notably, the crossover frequency shows little dependence on crosslinker concentration in the range $0.42 \leq c_x \leq 2.8$~wt\%, suggesting that the estimated relaxation time is robust across different network structures.
We thus conclude that the plateau length observed experimentally in Ref.~\cite{Mai2020} is consistent with our theoretical prediction based on Eq.~(\ref{eq:S:x_LS}).
This agreement supports the interpretation that the observed strain plateau ahead of the crack tip originates from viscoelastic relaxation rather than from microscopic fracture processes such as those described by cohesive-zone models.

\bibliography{bibliography}